\documentclass[a4paper,11pt]{article}
\usepackage{pos}
\usepackage{soul}

\title{Polarization measurements of the Crab Pulsar with POLAR}
 \ShortTitle{POLAR Crab Polarization}

\author*[1,2]{Hancheng Li}

\affiliation[1]{University of Geneva, Department of Astronomy, 16, Chemin d'Ecogia, 1290 Versoix, Switzerland}
\affiliation[2]{Key Laboratory for Particle Astrophysics, Institute of High Energy Physics, Beijing 100049, China}

\forColl{POLAR\footnote{\url{https://www.astro.unige.ch/polar/collaboration} }}
\emailAdd{Hancheng.Li@unige.ch}

\abstract{
	POLAR is a dedicated Gamma-Ray Burst polarimeter making use of Compton-scattering which took data from the second Chinese spacelab, the Tiangong-2 from September 2016 to April 2017. It has a wide Field of View of $\sim6$ steradians and an effective area of $\sim400\ cm^2$ at 300 keV. These features make it one of the most sensitive instruments in its energy range (15-500 keV), and therefore capable of almost continuously monitoring persistent sources such as pulsars. Significant folded pulsation from both PSR B0531+21 (the Crab Pulsar) and PSR B1509-58 has been observed. Observations of the Crab Pulsar with POLAR have previously been used for phase-resolved spectroscopy of the Crab Pulsar to calibrate the instrumental responses of POLAR. In this work, we investigate a polarimetric joint-fitting method for observations of the Crab Pulsar with POLAR. Unlike a GRB observation with POLAR, the observations of the Crab Pulsar are complicated by multiple observational datasets during which the polarization plane rotates as well. So before fitting, we have to correct the modulation curves under different datasets, by taking into account the rotations of the Crab Pulsar’s relative position in the detctor’s local coordinate, and the changes of detector response in different datasets. Despite these difficulties and the low signal to background for such sources constraining, polarization measurements were possible with the POLAR data. We will present the methodology briefly, which could be applied to any wide FoV polarimeter, and polarization results of the Crab pulsar with POLAR. Finally, the inferred ability of pulsar detection with POLAR-2 (the successor of POLAR) will also be discussed.
}

\FullConference{37$^{\rm{th}}$ International Cosmic Ray Conference (ICRC 2021)\\
		July 12th -- 23rd, 2021\\
		Online -- Berlin, Germany}

\begin{document}
\maketitle

\section{Introduction}\label{sec:Introduction}

The Crab pulsar is one of the most luminous X-ray sources in the sky, and thus it has been widely studied in the past decades. Many physical models~\cite[and references therein]{Harding2017} have been formed by the studies of the particle origin, acceleration mechanism, and the geometry of the radiation region, etc. Even though, there are still lots of unsolved mysteries. And thus the studies of the Crab pulsar remain in the frontier of astrophysics. As a new parameter, polarization is recognized as a unique probe to answer some of these questions. 

The POLAR detector is a dedicated wide FoV polarimeter of Gamma-Ray Bursts~\cite{Produit2018}. POLAR has no fixed-point observation capability, so its visibility of pulsars varies with its periodic motion in the Earth orbit, the period of which is determined by Tiangong-2 space lab with an average duration of $\sim$90 minutes. So each circle when pulsars pass by POLAR's Field of View (FoV), we can accumulate pulse signals in the observation data. This however led the analysis complicated by multiple observational parameters.

\section{Pulsar observation}\label{sec:Pulsars}

\begin{figure}[!htbp]
	\begin{center}
		\includegraphics[width=7.5cm,height=6cm]{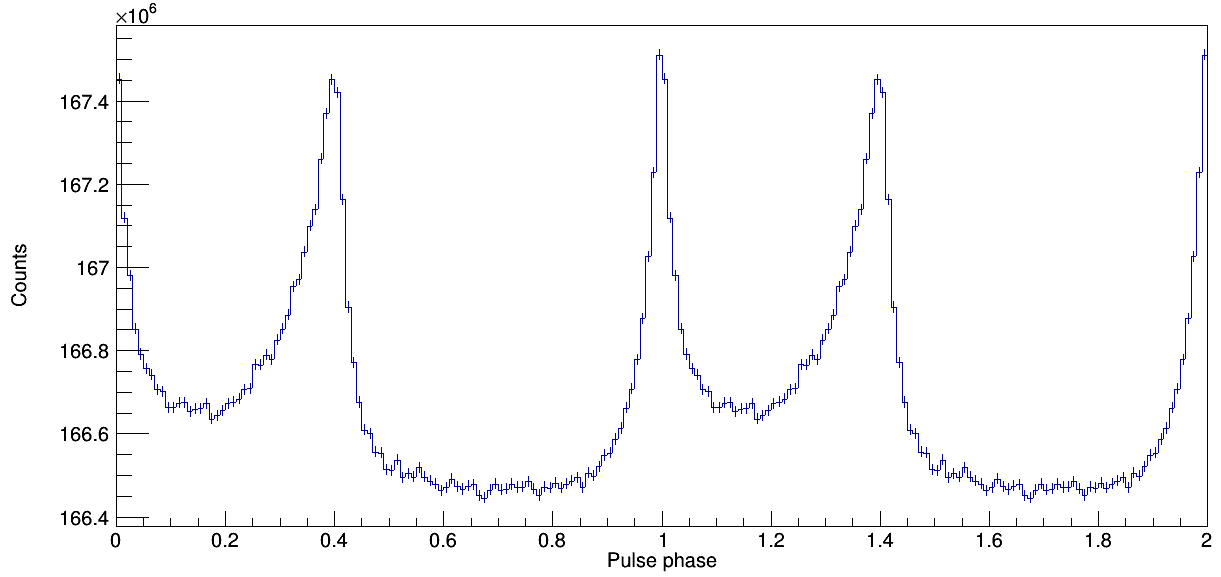}
		\includegraphics[width=7cm,height=6cm]{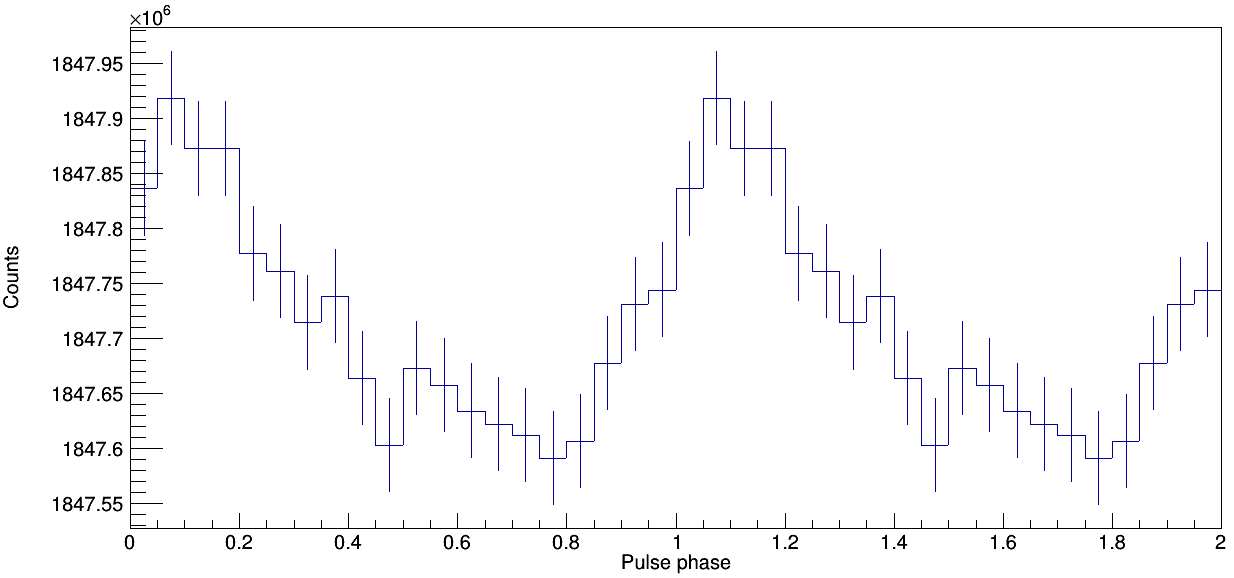}
		\caption{Pulse profiles of confirmed pulsars in POLAR data. The left plot is the pulse profile of the Crab pulsar, where phase intervals of P1, P2 and background are defined by 0.8$\sim$1.2, 0.2$\sim$0.6 and 0.6$\sim$0.8 respectively. The right plot is the pulse profile of PSR B1509-58.}
		\label{fig:POLAR_detected_psr}
	\end{center}
\end{figure}

At present, we have confirmed detections of the Crab pulsar and PSR B1509-58 in POLAR data. The pulse profiles are shown in Figure~\ref{fig:POLAR_detected_psr}. The timing analysis is based on~\cite{Hancheng2019}. And the best periodical parameters are shown in Table~\ref{tab:timing_parameters}. We have searched for other pulsars in POLAR data, but not confirmed yet. This work will be continued in the future. From the pulse profiles of the detected pulsars, we can see that the Signal-to-Noise Ratio (SNR) of Crab pulsar is much higher than that of the PSR B1509-58. Considering the statistical significance, our further studies are focused on the Crab pulsar.

\begin{table}[!ht]
	\centering
	\caption{The best periodical parameters of the Crab pulsar and the PSR B1509-58}\label{tab:timing_parameters}
	\begin{tabular}{l l c c}
		\hline
		&	Parameters				&	Crab pulsar			&	B1509-58	\\
		\hline
		&	$t_0$\ (MJD)			& 57697.040344079745	& 55336.0 \\
		&	$f_0$\ (Hz)				& 29.6484272934(4)		& 6.59709206418\\
		&	$f_1$\ (Hz\,s$^{-1}$)	& -3.689865(1)E-10		& -6.6531338E-11\\
		&	$f_2$\ (Hz\,s$^{-2}$)	& 1.16(1)E-20			& 1.8948E-21\\
		&	$f_3$\ (Hz\,s$^{-3}$)	& 3.4(3)E-28			& 0.0\\
		\hline
	\end{tabular}
\end{table}

\section{Previous works on the Crab pulsar}\label{sec:Previous}

This section we briefly introduce the previous work that we have done on the Crab pulsar, they served as calibrations of the detector, and thus laid the groundwork for the polarimetry.

\subsection{Simulation verification}\label{subsec:Simulation}

\begin{figure}[!htbp]
	\centering
	\includegraphics[width=9cm,height=7cm]{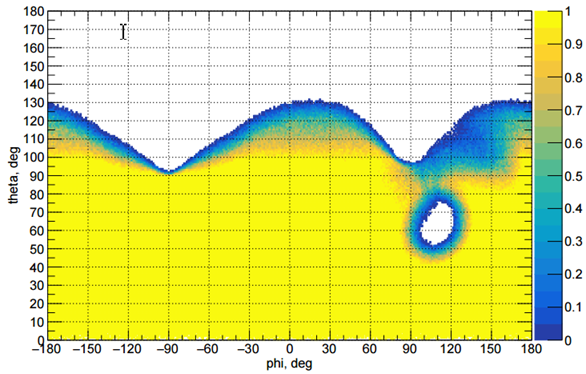}
	\caption{Simulation results of the visible efficiency as a function of incident angle. The missing efficiency at the position of about $\theta$ = 65$^{\circ}$ and $\phi$ = 100$^{\circ}$ is caused by shielding of the nearby antenna on Tiangong-2 platform.}
	\label{fig:crab_visible_eff}
\end{figure}

Some analyses on observation data has been performed in previous paper~\cite{Hancheng2017}. So here in this paper we introduce more on the simulation part. The factors that affect the observed counts to the Crab pulsar (\textit{D}) mainly include average detection efficiency $\eta$, projected geometric area $A(\theta,\phi)$, visibility efficiency $V(\theta,\phi)$ and spectrum $f(E)$. It can be expressed as:

\begin{equation}\label{eq:crab_detection_eff}
	D(\theta,\phi) = \eta\ A(\theta,\phi)V(\theta,\phi)\int_{E_{\rm min}}^{E_{\rm max}} f(E)dE \ ,
\end{equation}

where $\eta$ is a constant of the detector, which can be obtained by comparing the observed and simulated counts of the source. $A(\theta,\phi)$ can be calculated based on the geometry of the detector. $V(\theta,\phi)$ is simulated by Geant4, as shown in Figure ~\ref{fig:crab_visible_eff}.

\begin{figure}[!htbp]
	\begin{center}
		\includegraphics[width=7cm,height=6cm]{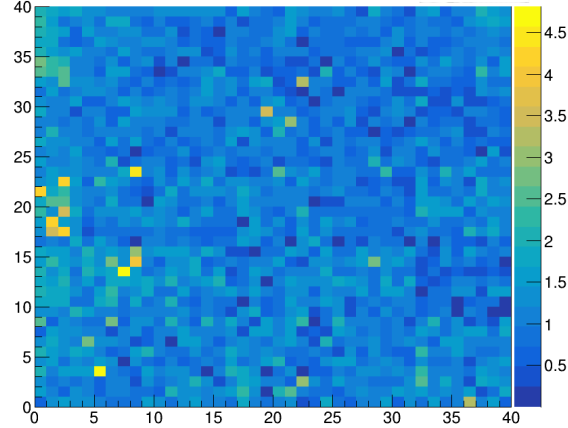}
		\includegraphics[width=7cm,height=6cm]{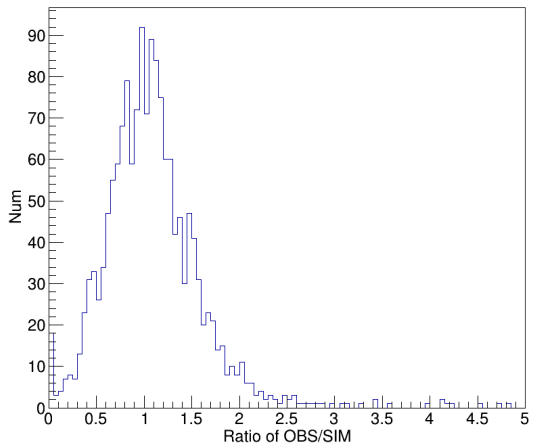}
		\caption{The average detection efficiency $\eta$ of 1600 detector channels. The value is obtained by (observed counts / predicted counts). The left plot shows the two-dimensional distribution of the average detection efficiency $\eta$ on the detector plane (40 x 40). The right figure is a one-dimensional histogram. It can be modelled as a Gaussian distribution with a mean of one.} 
		\label{fig:detetion_eff}
	\end{center}
\end{figure}

The input spectrum of the Crab pulsar is adapted from~\cite{Kuiper2001}:

\begin{equation}\label{eq:crab}
	f(E) = 726~E^{-1.276}~e^{-0.074~(\ln{E})^2} + 1464~E^{-1.165}~e^{-0.159~(\ln{E})^2} + 2021~E^{-2.022}\ ,
\end{equation}

where $f(E)$ is in units of ${\rm photons~keV^{-1}~m^{-2}~s^{-1}}$. We use this spectrum, together the real exposure time of the Crab pulsar to simulate the observation. Then the predicted counts of 1600 channels of the detector are obtained. The $\eta$ can be calculated by (observed counts / predicted counts), the results are shown in Figure~\ref{fig:detetion_eff}. The $\eta$ distribution can be modelled as a Gaussian distribution with a mean of $\sim$1. Therefore we can say $\eta$ is averagely equal to one in our analysis.

\begin{figure}[!htbp]
	\begin{center}
		\includegraphics[width=7cm,height=6cm]{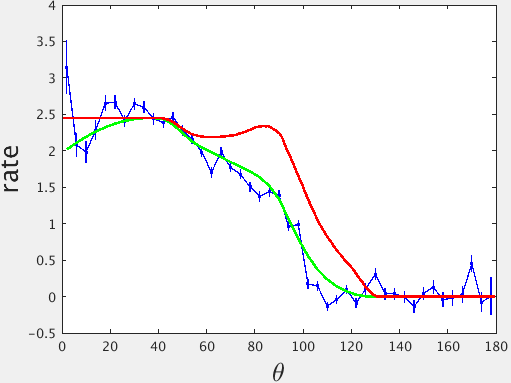}
		\includegraphics[width=7cm,height=6cm]{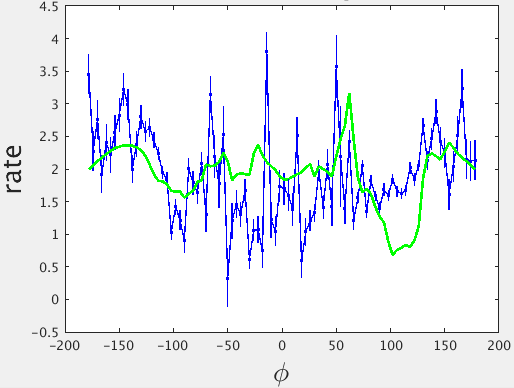}
		\caption{The count rate of the Crab pulsar from both simulation and observation. The left plot shows the count rate as a function of $\theta$. The right plot shows the distribution of count rate as a function of $\phi$. The blue curve in the plots are real observed values, the red line is integrated from $V(\theta,\phi)\int_{E_{\rm min}}^{E_{\rm max}} f(E)dE$, and the green lines are integrated from $A(\theta,\phi)V(\theta,\phi)\int_{E_{\rm min}}^{E_{\rm max}} f(E)dE$.}
		\label{fig:Crab_rate_fit}
	\end{center}
\end{figure}

The comparison between the prediction of~\ref{eq:crab_detection_eff} and the real count rate along with incident $\theta$ and $\phi$, are shown in Figure \ref{fig:Crab_rate_fit}. It can be seen that the predicted curves has a good fit to the real ones, which shows that the simulation is in good agreement with the real observation. Therefore the Geant4 model of the detector is adequate for providing responses to scientific studies.

\subsection{Phase-resolved spectroscopy}\label{subsec:spectroscopy}

\begin{figure}[!htbp]
	\begin{center}
		\includegraphics[width=11cm,height=8cm]{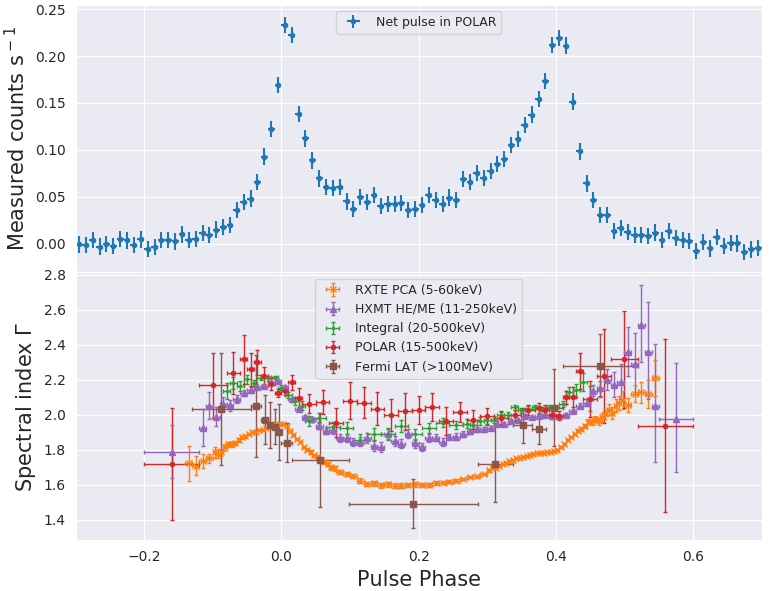}
		\caption{The spectral indices v.s. pulse phase of the Crab pulsar~\cite{Hancheng2019}. The top panel shows the net pulsed count rate of the Crab pulsar after background subtraction (background count rate is 26.88 counts/second), where the total exposure time is 1432349 seconds. The bottom panel shows the best fitted spectral index $\Gamma$ and its 1$\sigma$ standard error as a function of pulse phase.}
		\label{fig:spectral}
	\end{center}
\end{figure}

Before proceeding with the polarization studies of the Crab pulsar using POLAR data, we decided to further check the accuracy of instrumental responses of all incoming directions by performing a phase-resolved spectroscopy analysis~\cite{Hancheng2019}. After extracting spectrum and joint-fitting to it, we obtain the spectral indices (single power-law) as a function of pulse phase of the Crab pulsar, which are shown in Figure~\ref{fig:spectral}. Overall, the distribution of the spectrum indices, showing an "reverse-S" pattern, is consistent with other results in the same energy range within the margin of errors. It can verify the results of other missions in the same energy range, and also plays an important role in calibration of the POLAR detector.

\section{Polarimetry}\label{sec:Polarimetry}

\begin{figure}[!htbp]
	\begin{center}
		\includegraphics[width=6cm,height=6.5cm]{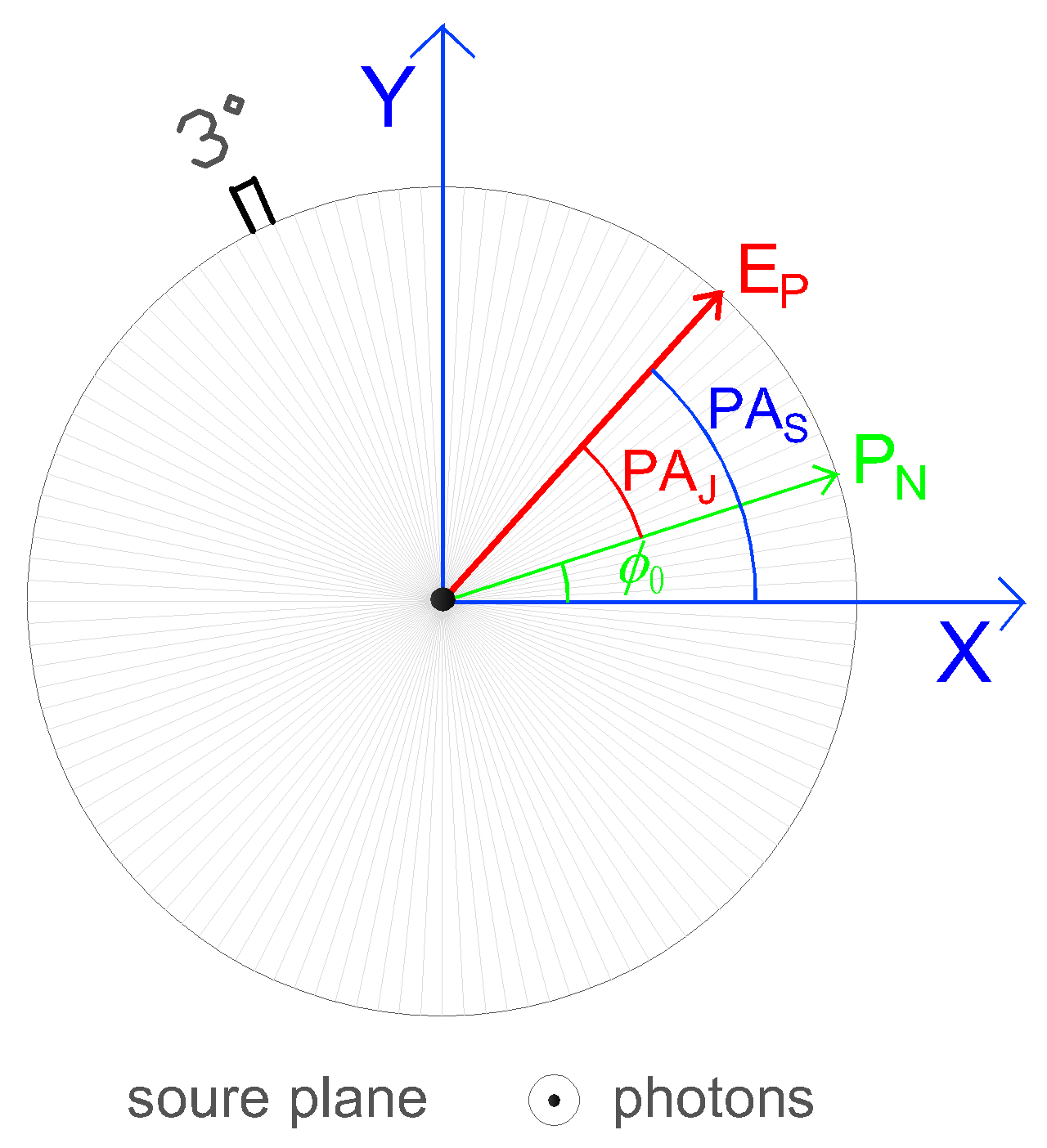}
		\quad
		\includegraphics[width=7cm,height=6cm]{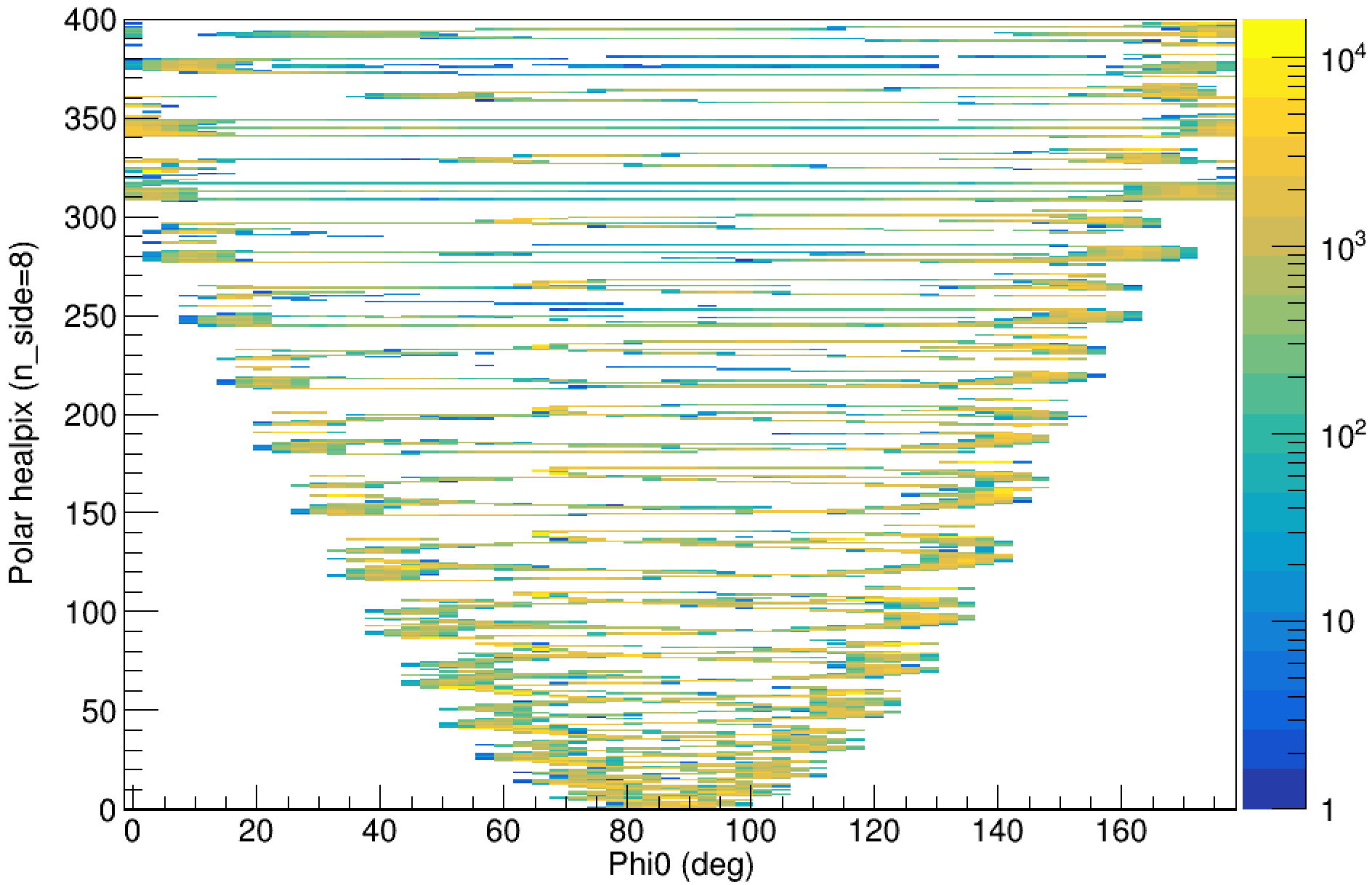}
		\caption{The rotation of the Crab pulsar. The left plot shows the plane that is perpendicular to the incident direction of the Crab pulsar. In this plane we have seen that the Crab pulsar (together with projection of the North Pole vector of the equatorial coordinate) is rotating. We define $\phi_0$ angle to describe this rotation. The right plot is the exposure time as a function of $\phi_0$ (X-axis) and incident angle (Y-axis, in HEALPIX numbering).}
		\label{fig:exposure_phi0}
	\end{center}
\end{figure}

From Figure~\ref{fig:exposure_phi0}, the left plot of which is a plane perpendicular to the incident direction of the Crab pulsar, we found that the polarization vector of the Crab pulsar (\textbf{$\rm E_P$}), and the projection of the North Pole vector of equatorial coordinate on the Polarization plane (or Source plane) (\textbf{$\rm P_N$}), are still co-rotating on that plane even at the same incident direction. That plane is just two-dimensional, and the source is three-dimensional, so even if the Crab pulsar is from one incident direction, it may still have rotation there. We define the angle of $\phi_0$ to describe such rotation. The right plot shows the exposure time of POLAR to Crab pulsar as a function of $\phi_0$ and incident angle.

Therefore, we develop a de-rotation method on the $\phi_0$, and finally we can obtain one total modulation curve for each incident direction. And we verified that this method is adequate for the reconstruction of polarization, including Polarization Angle (PA) and Polarization Degree(PD). Limited by the paper length, we skip the technical details about the de-rotation method, but it will be refered to ~\cite{Hancheng2021}.

Then we fit the modulation curves of every sky directions with corresponding simulated modulation curves database jointly, the $\chi^2$ statistic is calculated by:

\begin{equation}\label{equ:chi2}
\chi^2=\sum_{i=1}^n \frac{(X_i-Y_i)^2}{\varepsilon_i^2+\sigma_i^2},
\end{equation}

where $X_i$ and $Y_i$ are the counts of $i$th bin in simulated and observed modulation curves respectively,  $\varepsilon_i$ and $\sigma_i$ are the corresponding uncertainties of $X_i$ and $Y_i$.

And finally, the preliminary polarization measurement results of the Crab pulsar by POLAR with the analysis method described above is shown in Table~\ref{tab:polarization}. Our polarization results agree with that of POGO+~\cite{Chauvin2017}. Particularly, our results suggest that the PA of averaged phase is quasi-parallel to the rotaion axis of the Crab pulsar. Note that in our analysis, the Nebula contribution (off pulse region) has been subtracted. Additionally, we are currently working on a Bayesian approach to get posterior on PD and PA, this allows to give deviation and upper/lower limits more precisely.

\begin{table}[!ht]
	\centering
	\caption{Polarization results of the Crab pulsar with POLAR (Definitions of P1/P2 is shown in Figure~\ref{fig:POLAR_detected_psr}).}\label{tab:polarization}
	\begin{tabular}{c c c c}
		\hline
		&	Phase range		&	PA ($\deg$)	&	PD (\%)	\\
		\hline
		&	Averaged phase (-Nebula)	& 120	& 17 \\
		&	P1				& 174		& 19\\
		&	P2	            & 81		& 23\\
		\hline
	\end{tabular}
\end{table}

\begin{figure}[!htbp]
	\begin{center}
		\includegraphics[width=10cm,height=7cm]{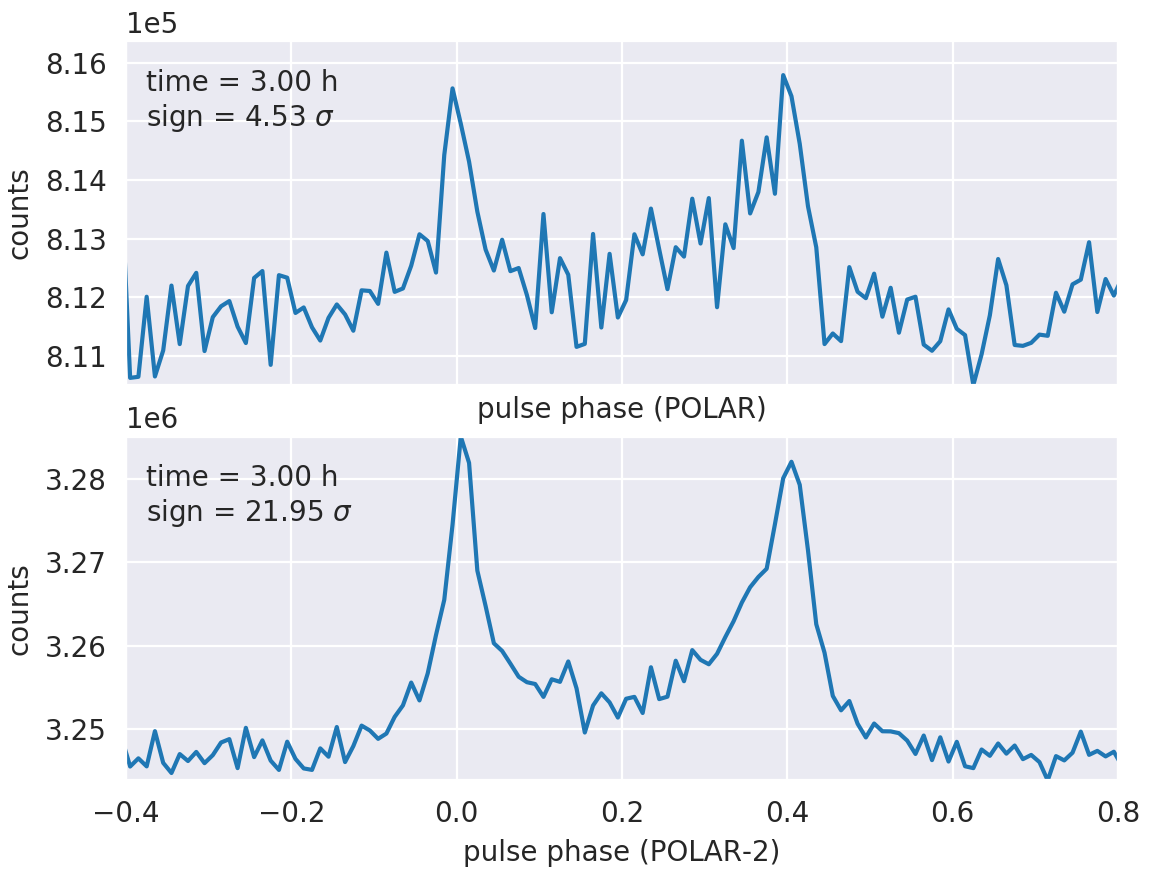}
		\caption{Three-hour-detection of the Crab pulsar with POLAR and POLAR-2}
		\label{fig:polar2}
	\end{center}
\end{figure}

POLAR-2, as a following experiment of POALR, has already been selected to be launched on the China Space Station around 2024, and it is now being constructed~\cite{Kole2021, NDA2021}. Our previous work has estimated the detection of Crab pulsar with POLAR-2~\cite{Hancheng2019}. Figure~\ref{fig:polar2} shows typical three-hour-detection with POLAR and POLAR-2. As one can see, the POLAR-2 has much more significant pulsation than POLAR. This will lead to all above analyses being worked out better by POLAR-2. 

\section{Conclusion}\label{sec:Conclusion}
POLAR is adequate for the polarimetry of pulsars. It found the Crab pulsar and the PSR 1509-58 for now, and measured the polarization of the Crab pulsar. The methodology could be adapted for other wide-FoV polarimeters. In the near future, POLAR-2 will be able to to find more pulsars, and contribute polarization measurements with much better precision. As for the Crab pulsar, we expect to do phase-resolved and energy-resolved polarimetry which will greatly help us further understanding the physics of such source, as well as its radiation emission mechanism and the geometry of the radiation region.

\section*{Full Authors List: POLAR Collaboration}


\scriptsize
\noindent
{Hancheng Li}$^{1,2,\ast}$,
{Nicolas Produit}$^{1}$,
{Merlin Kole}$^{3}$,
{Jian-Chao Sun}$^{2,4}$,
{Ming-Yu Ge}$^{2}$,
{Yuan-Hao Wang}$^{2,4}$,
{Nicolas De Angelis}$^{3}$,
{Neal Gauvin}$^{1}$,
{Wojtek Hajdas}$^{5}$,
{Johannes Hulsman}$^{3}$,
{Zheng-Heng Li}$^{2,4}$,
{Li-Ming Song}$^{2,4}$,
{Teresa Tymieniecka}$^{6}$,
{Bo-Bing Wu}$^{2,4}$,
{Xin Wu}$^{3}$,
{Shao-Lin Xiong}$^{2}$,
{Yong-Jie Zhang}$^{2}$,
{Yi Zhao}$^{2,7}$,
{Shi-Jie Zheng}$^{2}$
and
{Shuang-Nan Zhang}$^{2,4}$ \\

\noindent
$^1${University of Geneva, Department of Astronomy, 16, Chemin d'Ecogia, 1290 Versoix, Switzerland} \\
$^2${Key Laboratory for Particle Astrophysics, Institute of High Energy Physics, Beijing 100049, China} \\
$^3${Department of Nuclear and Particle Physics, University of Geneva, 24 Quai Ernest-Ansermet, 1205 Geneva, Switzerland} \\
$^4${University of Chinese Academy of Sciences, Beijing 100049, China} \\
$^5${Paul Scherrer Institut, 5232, Villigen, Switzerland} \\
$^6${National Centre for Nuclear Research, ul. A. Soltana 7, 05-400 Otwock, Swierk, Poland} \\
$^7${Department of Astronomy, Beijing Normal University, Beijing, 100875, China}

\end{document}